\documentclass[%
 aip,
 amsmath,amssymb,
 reprint,%
]{revtex4-1}
\usepackage{xcolor}
\usepackage{graphicx}% Include figure files
\usepackage{dcolumn}% Align table columns on decimal point
\usepackage{bm}% bold math
\usepackage[utf8]{inputenc}
\usepackage[T1]{fontenc}
\usepackage{mathptmx}
\usepackage{etoolbox}

\makeatletter
\def\@email#1#2{%
 \endgroup
 \patchcmd{\titleblock@produce}
  {\frontmatter@RRAPformat}
  {\frontmatter@RRAPformat{\produce@RRAP{*#1\href{mailto:#2}{#2}}}\frontmatter@RRAPformat}
  {}{}
}%
\makeatother
\begin{document}

\preprint{AIP/123-QED}

\title{Josephson Traveling Wave Parametric Amplifiers\\ with Plasma oscillation phase-matching}
% Force line breaks with \\
\author{E. Rizvanov}
 
\email{emil.rizvanov@fmph.uniba.sk}

\author{S. Kern}%

\author{P. Neilinger}
 
\altaffiliation{ 
Institute of Physics, Slovak Academy of Sciences,
		D\'{u}bravsk\'{a} cesta, Bratislava, Slovakia
}%

\author{M. Grajcar}
 
\altaffiliation{ 
Institute of Physics, Slovak Academy of Sciences,
		D\'{u}bravsk\'{a} cesta, Bratislava, Slovakia
}%

\affiliation{ 
Department of Experimental
		Physics, Comenius University, SK-84248 Bratislava, Slovakia
}%

\date{\today}% It is always \today, today,

\begin{abstract}
High gain and large bandwidth of traveling-wave parametric amplifier exploiting the nonlinearity of Josephson Junctions can be achieved by fulfilling the so-called phase-matching condition. This condition is usually addressed by placing resonant structures along the waveguide or by periodic modulations of its parameters, creating gaps in the waveguide's dispersion. Here, we propose to employ the Josephson junctions, which constitute the centerline of the amplifier, as resonant elements for phase matching.
By numerical simulations in JoSIM (and WRspice) software, we show that Josephson plasma oscillations can be utilized to create  wave vector mismatch sufficient for phase matching as well as to prevent the conversion of the pump energy to higher harmonics. The proposed TWPA design has a gain of 15 dB and a 3 GHz bandwidth, which is comparable to the state-of-the-art TWPAs.

\end{abstract}

\maketitle

\section{\label{sec:Intro}Introduction}

The Josephson Traveling Wave Parametric Amplifier (JTWPA) is a superconducting device that exploits the nonlinearity of Josephson junctions embedded in a transmission line (TL) to amplify weak microwave signals with minimal added noise. The principle of operation is based on the interaction of the signal wave co-propagating with a strong driving wave, called the pump, via a nonlinear mixing process known as parametric amplification \cite{Cullen1960TheoryOT, PhysRevA.39.2519, First_JTWPA, PhysRevB.87.144301}.
JTWPAs, as a practically lossless medium\cite{Devoret}, can amplify signals with noise performance approaching the quantum limit, making them ideal for quantum computing and sensitive measurement applications\cite{Macklin, Devoret}.
Among the biggest challenges in TWPAs development are the mitigation of unwanted nonlinear processes, like nonlinear phase shifts and the higher-harmonic generation \cite{9134828}. The nonlinear phase shift arises as the strong pump propagates at a higher phase velocity, which leads to the deterioration of the amplifier's performance. Conversely, the generated higher harmonics cause a leakage of the pump power and its depletion \cite{Dixon}. 

These challenges were addressed by so-called dispersion engineering - adjusting the dispersion relation of the TL to compensate the nonlinear phase-shift of the pump and to avoid the higher-harmonic generation.
Several designs of JTWPAs have been proposed and studied:
 Photonic-Crystal design \cite{Planat2019}, Coupled Asymmetric SQUIDs chain \cite{Samolov, Ranadive_2022}, SQUID Chain  Design \cite{Zorin2016,Zorin2017,Dixon,Gaydamachenko2022,Kissling2023,Nilsson2023,Peatain2021,levochkina2024}, Resonator Phase Matching \cite{OBrien2014,White2015,Kissling2023,Peng_2022,Peng_FM, Macklin},  and Bi-SQUID Based Design \cite{BiSQUID}.

In this paper, we propose a simple design of JTWPA, enabling phase-matching and preventing the generations of higher harmonics.

We have chosen the 3WM nonlinear processes for amplification, since they have distinct advantages relative to four-wave mixing (4WM) \cite{Zorin2016,Vissers,Dixon}.  Namely lower pump power, reduced noise, broader bandwidth, and more stable operation. These benefits make 3WM particularly suitable for applications in quantum information processing and in sensitive, low-noise amplification. Although 4WM offers advantages such as higher potential gain and more flexible frequency conversion, our 3WM JTWPA, with plasma oscillation phase matching, provides a high 20 dB gain, which is sufficient for preamplifying signals for commercial HEMT amplifiers.

First, we provide a straightforward method to analyze transmission lines with resonant elements and their utilization in dispersion engineering. Then, we describe a metawaveguide where the Josephson junctions provide both the nonlinear inductance and the resonant characteristics at the junction's plasma frequency. The plasma frequency is determined by the junction's own capacitance and by an additional, designed one. The feasibility of this design, as estimated from basic TWPA theory, is confirmed by numerical simulation of a TL consisting of 2000 JJs. The results of the simulations provide further insight into the influence of plasma oscillations on nonlinear processes as well as harmonics generation and pump depletion.

Numerical calculations enable us to consider higher harmonics\cite{Dixon} and other nonlinear processes beyond the reach of analytical methods. For simulations we utilized recently developed software, JoSIM\cite{Delport2019, JoSIMSite}, which employs  modern numerical methods to analyze Josephson junction circuits. We compare the performance of this software with the most renowned one, WRspice\cite{133816,Spice_cite}, which is widely used in the design and analysis of numerous amplifiers\cite{Dixon, Peatain2021, Kissling2023, Gaydamachenko2022, Peng_2022, Peng_FM}. We verified that our device performs excellently in both softwares, although the magnitude of the gain and the ripples are slightly different, the overall shape of the gain profile is the same.

\section{\label{sec:Intro} CIRCUIT DESCRIPTION}

\subsection{\label{sec:Gain} Gain of the amplifier}

\label{s:Gain}

The processes in TWPAs are commonly explained within the coupled mode
equations (CME) describing mixing between the pump, signal, and idler with respective angular frequencies $\omega_p,\omega_s,\omega_i$ satisfying $\omega_p=\omega_s+\omega_i$.
In general, wave vectors corresponding to these modes exhibit dispersion, leading to the difference in the wave vectors $\Delta k=k_p-k_s-k_i$.
The CME result for the gain of TWPA with length $L$ under the three-wave-mixing (3WM) is
\begin{equation}
    G(L) = \cosh^2\left(gL\right)+\left(\frac{\beta}{2g}\right)^2\sinh^2\left(gL\right)
    \label{eq:gain}
\end{equation}
with the gain per unit length
\begin{equation}
    g = \sqrt{\frac{k_sk_i}{4}\left(\frac{I_pI_d}{I_d^2+I_\ast^2}\right)^2-\frac{\beta^2}{4}},
    \label{eq:g}
\end{equation}
and the effective nonlinear phase shift per unit length
\begin{equation}
    \beta = \Delta k\left(1+\frac{I_p^2}{4(I_d^2+I_\ast^2)}\right)-\frac{I_p^2}{8(I_d^2+I_\ast^2)}k_p.
    \label{eq:b}
\end{equation}
Here, $I_\ast = \sqrt{2}I_c$, where $I_c$ is critical current of Josephson junction, $I_p$ and $I_d$ are the amplitudes of the pump and DC current, respectively.  Adjusting the dispersion $k(\omega)$ to minimize  $\beta$ leads to higher gain. Moreover, $\beta=0$ results in a much larger bandwidth and maximal gain, which scales exponentially with the TWPA's length, in contrast to the quadratic gain obtained from linear dispersion.

\subsection{\label{sec:Gain} Dispersion relation}

\begin{figure}
	\includegraphics[width=8.6cm]{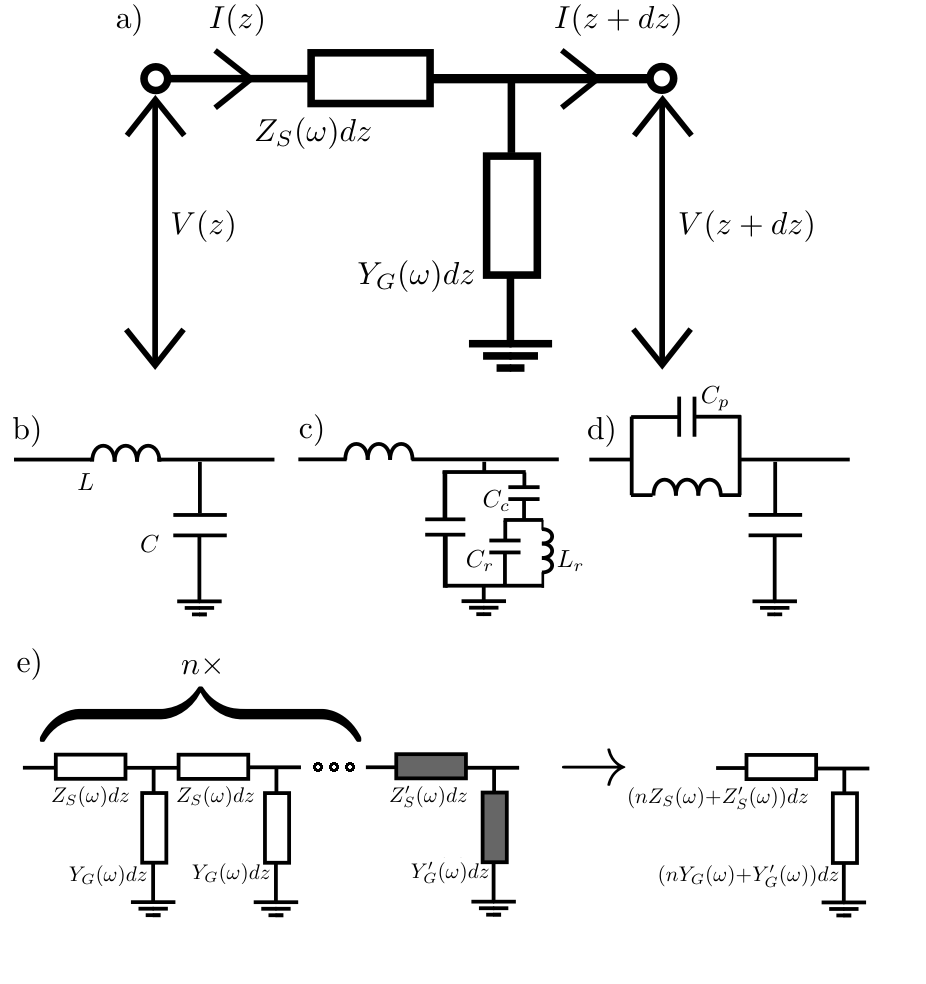}
	\caption{ Scheme of unit cell of: a) resonator embedded in transmission line; b) simple lossless TL; c)  LC resonator; d)  LC resonator embedded in series to transmission line; e) resonator every nth element.}
	\label{fig:Scheme}
\end{figure}

To estimate the dispersion relation $k(\omega)$ in a transmission line with various resonant elements, the standard  TL model \cite{Pozar:882338} is modified (Fig. \ref{fig:Scheme}b).  The original series inductance and ground capacitance are replaced with a general series impedance  $Z_S(\omega)dz$ and ground admittance $Y_G(\omega)dz$ in harmonic representation at frequency $\omega$ (see Fig. \ref{fig:Scheme}a). Writing Kirchhoff's laws for the current and the voltage drop between points $z$ and $z+dz$
\begin{equation}
\begin{aligned}
    V(z,\omega)-V(z+dz,\omega) &= Z_S(\omega)dz I(z,\omega)\\
    I(z,\omega)-I(z+dz,\omega) &= Y_G(\omega)dz V(z+dz,\omega)
    \label{eq:Kirchhoff_law}
\end{aligned}
\end{equation}
we can derive the wave equation
\begin{equation}
    \frac{d^2I(z,\omega)}{dz^2} = -\frac{\omega^2}{v_p^2}I(z,\omega),
\end{equation}
where one can identify the frequency dependant phase velocity 
\begin{equation}
    v_p(\omega) = i\frac{\omega}{\sqrt{Z_S(\omega)Y_G(\omega)}},
    \label{eq:phase_velocity}
\end{equation}
which determines the TL dispersion $k(\omega)=\omega/v_p(\omega)$, where $k(\omega)$ is the wave vector of propagating wave with frequency $\omega$.
Indeed, for simple lossless TL at Fig. \ref{fig:Scheme}b we obtain $Z_Sdz = i\omega L$ and $Y_Gdz = i\omega C$, giving the well known result for the constant phase velocity $v_p=dz/\sqrt{LC}$.

To modify the dispersion $k(\omega)$, various resonant elements can be introduced via the frequency dependence of the parameters $Z_S(\omega)$ and $Y_G(\omega)$. Let us start with an LC-resonator proposed in Refs.~\onlinecite{white2015traveling, OBrien2014}  as depicted in Fig. \ref{fig:Scheme}c. The resulting dispersion relation of such TL 
\begin{equation}
    k(\omega) = \omega\frac{\sqrt{LC}}{dz}\sqrt{1+\frac{C_c}{C}\frac{1-\omega^2/\omega_r^2}{1-\omega^2/\omega_0}},
\end{equation}
results in a phase shift near the frequency $\omega_0=\omega_r\omega_c/\sqrt{\omega_r^2+\omega_c^2}$, where $\omega_r=1/\sqrt{L_rC_r}$, and $\omega_c=1/\sqrt{L_rC_c}$. This phase shift was utilized as the so-called resonant phase-matching in many recent TWPA designs \cite{OBrien2014,White2015,Kissling2023,Peng_2022,Peng_FM, Macklin}.

Alternatively, the resonant element can be incorporated into the series impedance $Z_S(\omega)$.
The simplest way is to connect a capacitor in parallel to the TL inductance as in Fig. \ref{fig:Scheme}d. This creates a phase shift across the whole bandwidth and is divergent at the resonant frequency:
\begin{equation}
    k(\omega) = \omega\frac{\sqrt{L_JC}}{dz}\sqrt{\frac{1}{1-\omega^2/\omega_p^2}},
\end{equation}
where $\omega_p=1/\sqrt{L_JC_J}$ is the plasma frequency of a single  junction. 
Above $\omega_p$  the wave vector is purely imaginary. The characteristic impedance of the waveguide defined as
\begin{equation}
    Z = \sqrt{\frac{Z_s}{Y_G}}
\end{equation}
determines the reflection coefficient $\Gamma = (Z_L-Z)/(Z_L+Z)$, which became unity above $\omega_p$. Here $Z_L$ is the impedance of the input lines, typically $50~\Omega$. The reflection of all tones above the plasma frequency avoids higher harmonic generation.

\begin{figure}
	\includegraphics[width=8.6cm]{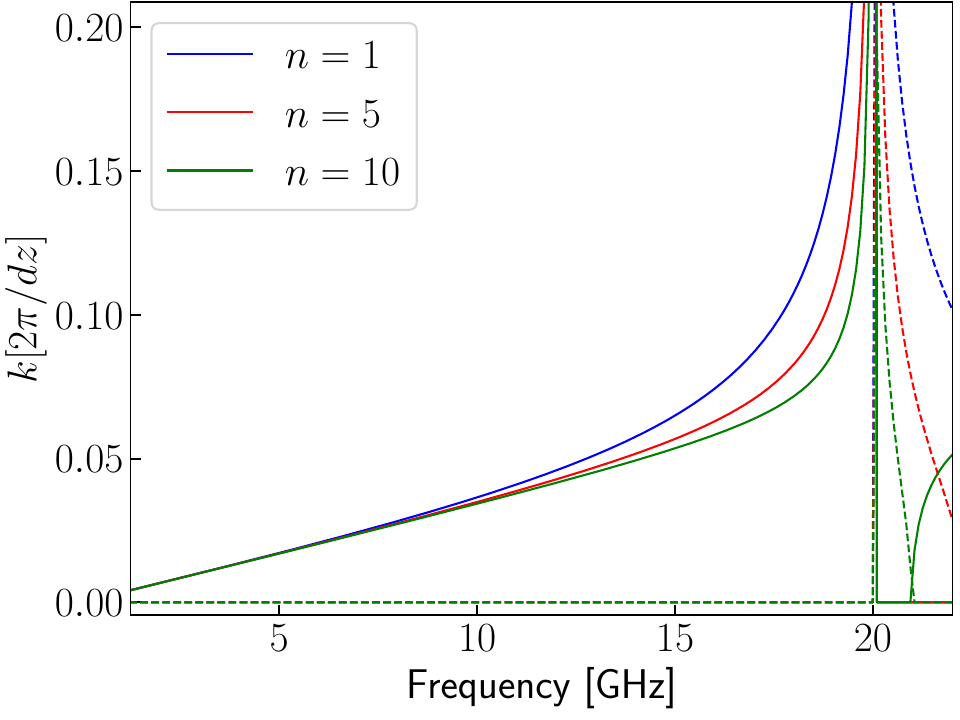}
	\caption{ Dispersion curves given by Eq. (\ref{eq:plas_dis}) for  $n=1,5,10$ (blue,  red and green  lines respectively). Solid and dashed lines are real and imaginary parts of $k$ respectively. The wave vector is normalized to $2\pi/dz$, where $dz$ is the distance between adjacent junctions as defined in Fig. \ref{fig:JJ_RPM_pic}. The dispersion curves diverge near 
 plasma frequency, which provides frequency dependent phase shift for phase matching. Their shape  can be controlled by a value of $n$. Above the plasma frequency, the wave vector becomes imaginary, which does not allow the propagation of waves above plasma frequency.} 
	\label{fig:reflect}
\end{figure}

For our amplifier to operate in the frequency range from $2$ \textrm{GHz} to $7$ \textrm{GHz} in 3-wave mixing, the pump ($f_{p}$) has to be set at about 10 \textrm{GHz}. Therefore, to suppress higher harmonics of the pump and to achieve phase matching, the frequency gap should be below 20 \textrm{GHz}. Thus, we choose critical current $I_{c}$ and the parallel capacitance $C_{p}$ such that
$1/2\pi\sqrt{L_JC_p} = \sqrt{I_c/2\pi\Phi_0C_p}\approx 20~$\textrm{GHz}. Since the capacitance $C_p\approx 400\ \textrm{fF}$ is determined by the size of the capacitor which is the largest element of the TWPA it is
convenient to place them  equidistantly into every $n$-th element of the TL as depicted in the  Fig. \ref{fig:Scheme}e.

\begin{figure}
	\centering  
	\includegraphics[width=8.6cm]{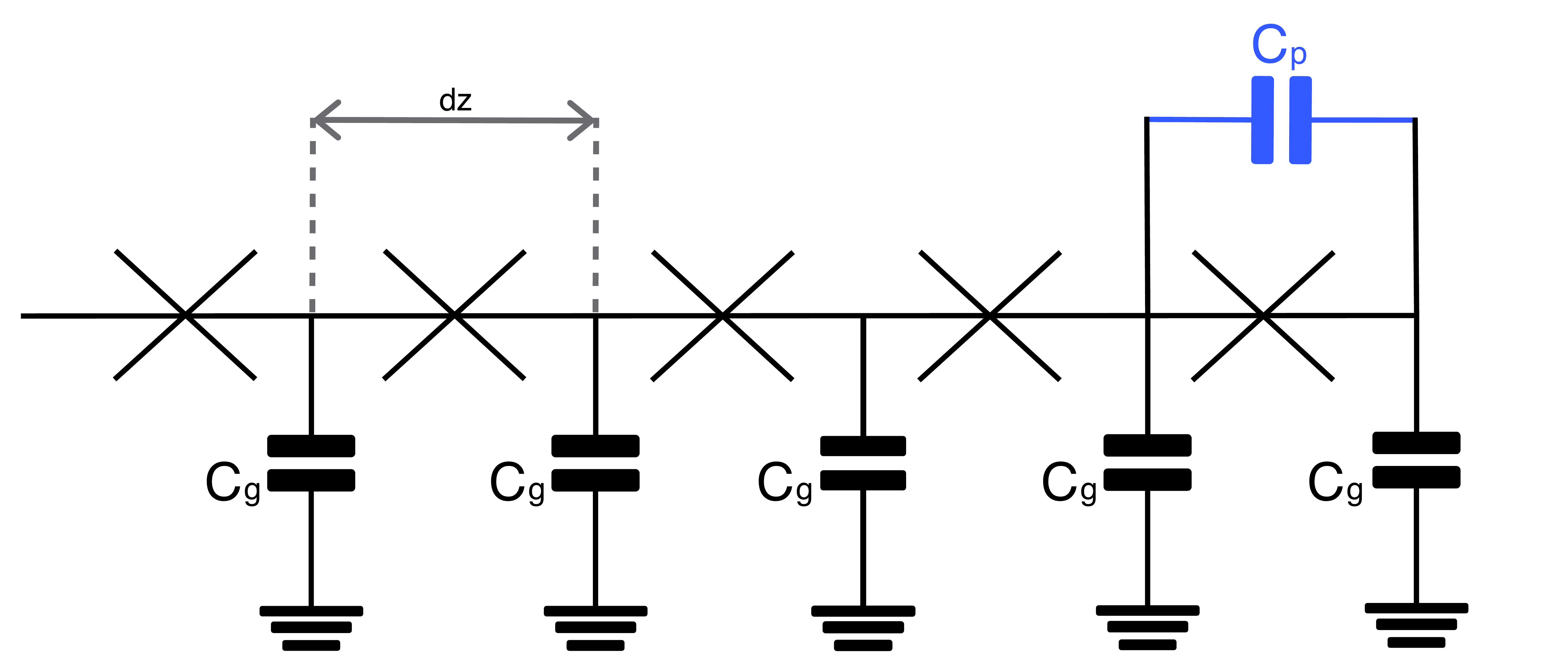}
       
	\caption{Schematic of the TWPA units cell. Capacitor $C_{p} = \textrm{394 fF}$  (blue) responsible for plasma frequency placed in parallel to every fifth ($n = 5$)   Josephson junction (black crosses). All Josephson junctions have the same RSCJ model:$     \;   I_{c} =\textrm{2 uA}, R = \text{550 } \Omega, C_{JJ} =\textrm{12 fF}$. All ground capacitances have equal values: $ C_{g} =  \textrm{71.5 fF}$. In simulations, the whole circuit contains $400$ unit cells, thus the total amount of junction is 2000.}
  \label{fig:JJ_RPM_pic}
	
\end{figure}

Such system can be analyzed by the introduced formalism, assuming the spacing between the resonators is small in comparison to the wavelength $ndz\ll\lambda$. Namely, it can be shown (see Appendix \ref{app:elem}) that the following model to first order in $dz$ reduces to $n$-times smaller influence of the resonant phase shift. We obtained following dispersion relation 
\begin{equation}
    k(\omega)=\omega\frac{\sqrt{L_JC}}{dz}\sqrt{\frac{n}{n+1}}\sqrt{1+\frac{1}{n}\frac{1}{1-\omega^2/\omega_p^2}}.
    \label{eq:plas_dis}
\end{equation}

We choose $n=5$, which leaves space for the large capacitor $C_{p} = 396 \textrm{ fF}$, and the distance between them is still less than the wavelength. For example, this capacitance could be provided by a miniature $Al_{2}O_{3}$ parallel capacitors. At a dielectric thickness of $d=10 \textrm{ nm}$, the capacitor area of $A=20\textrm{ um}^{2}$ would be sufficient. Capacitors with similar capacitance per area were fabricated
using hexagonal boron nitride \cite{capacitor}.

\begin{figure}
\includegraphics[width=8.6cm]{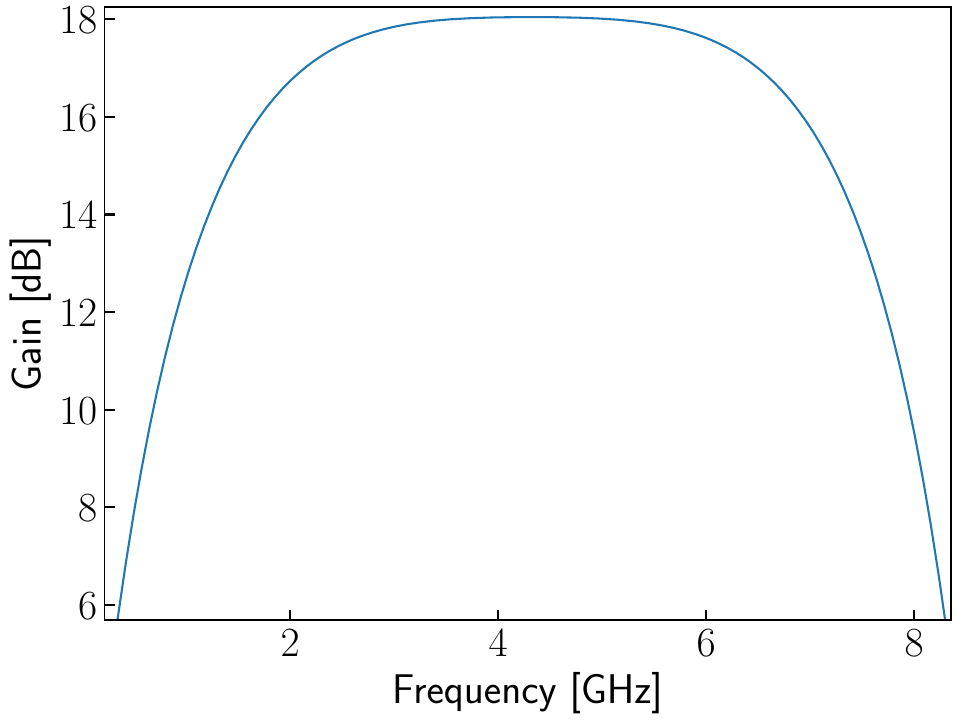}
	\caption{Gain profile obtained by CME (see Eqs.~(\ref{eq:gain}, \ref{eq:g}, \ref{eq:b}, \ref{eq:plas_dis})) for $I_{d} =0.4I_{c},I_{p} \approx 0.4I_{c}, f_{p}=8.64\textrm{ GHz}, I_{c} = 2 \textrm{ uA}$. Here $I_{d}$, $I_{p}$, $f_{p}$, $I_{c}$ are amplitudes of dc current, pump current, pump frequency, critical current, respectively. }
	\label{fig:Gain_CE}
\end{figure}

The dispersion relation for the introduced design (see Fig.~\ref{fig:Scheme}) is depicted in Fig.~\ref{fig:reflect}.
Substituting the dispersion relation Eq.~(\ref{eq:plas_dis}) to Eqs.~(\ref{eq:gain}, \ref{eq:g}, \ref{eq:b}) we obtained the gain profile depicted in Fig.~\ref{fig:Gain_CE}. The gain is more than 15 \textrm{dB} with a desired flat profile from $3\textrm{ GHz} $ to $6\textrm{ GHz}$.

\begin{figure}
\includegraphics[width=8.6cm]{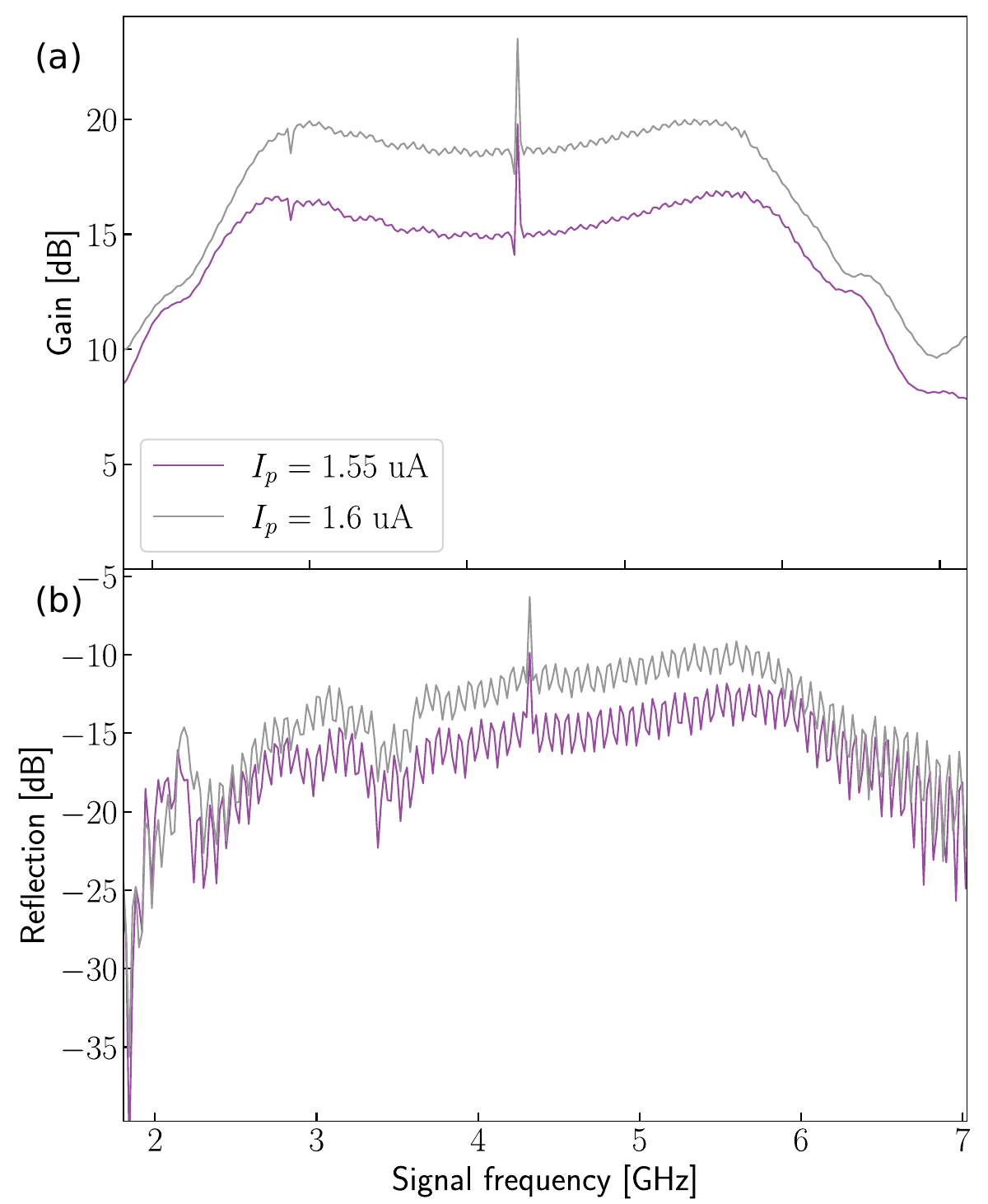}
	%\label{fig:Gain}
	\caption{  Gain (a) and reflection (b) simulated in JoSIM for $I_{s} = 0.01\textrm{ uA}$, $I_{d} =0.8\textrm{ uA}, f_{p}=8.64\textrm{ GHz }$, and two values of pump amplitudes: $I_{p} = 1.55\textrm{ uA}$ (purple),  $I_{p} = 1.6\textrm{ uA}$ (grey).  $I_p$, $I_s$, and $I_d$ denote the amplitudes of the current waves in the TWPA. These amplitudes are half of the values supplied by an impedance-matched current source, as described in Ref. \onlinecite{Zorin2017}. Increasing the pump amplitude can enhance the gain, but it will also increase reflections, leading to ripples. }
 \label{fig:Gain}
\end{figure}

\section{\label{sec:Sim}Simulation}

To verify the feasibility of the proposed design and study processes not covered by the CME, we utilized the recently developed numerical simulator JoSIM \cite{Delport2019,JoSIMSite} to model JTWPA. We verified this software by comparing the results with well-known WRspice \cite{133816,Spice_cite} (see Appendix \ref{app:comp}).

The simulation setup is the same as in Ref.~\onlinecite{Gaydamachenko2022}. The studied scheme is in Fig. \ref{fig:JJ_RPM_pic} consisting of 2000 JJs.  From the estimated phase velocity $v_p\sim dz/\sqrt{L_JC_g}$, where $dz$ is distance between JJs, the travelling time can be estimated $t = 2000dz/v_p\approx6.6~\textrm{ns}$. In order to exclude transient effects\cite{Dixon_thesis}, we saved the input and the output currents and voltages at the first and the last node from $25 \textrm{ ns}$ to $75 \textrm{ ns}$.
We  checked that our simulations are in the steady state. 
For example, in Ref. \onlinecite{Gaydamachenko2022} the steady state was  achieved in $10  \textrm{ ns}$.  
The simulation time step was set to 0.1 ps, which is significantly smaller than the one used in Ref. \onlinecite{Gaydamachenko2022}. We have verified that the results of the simulations remain consistent with smaller time steps.

This way, we obtained overall gain  and power in each node n \cite{Pozar:882338, Dixon, Gaydamachenko2022}. The obtained gain from the simulations (see Fig. \ref{fig:Gain}) is higher than $15 \textrm{ dB}$ similarly to the CME result in Fig. \ref{fig:Gain_CE} with  reflection coefficient not exceeding $-10 \textrm{ dB}$ in the range from $3$ to $6 \textrm{ GHz}$.

\begin{figure}

	\includegraphics[width=8.6cm]{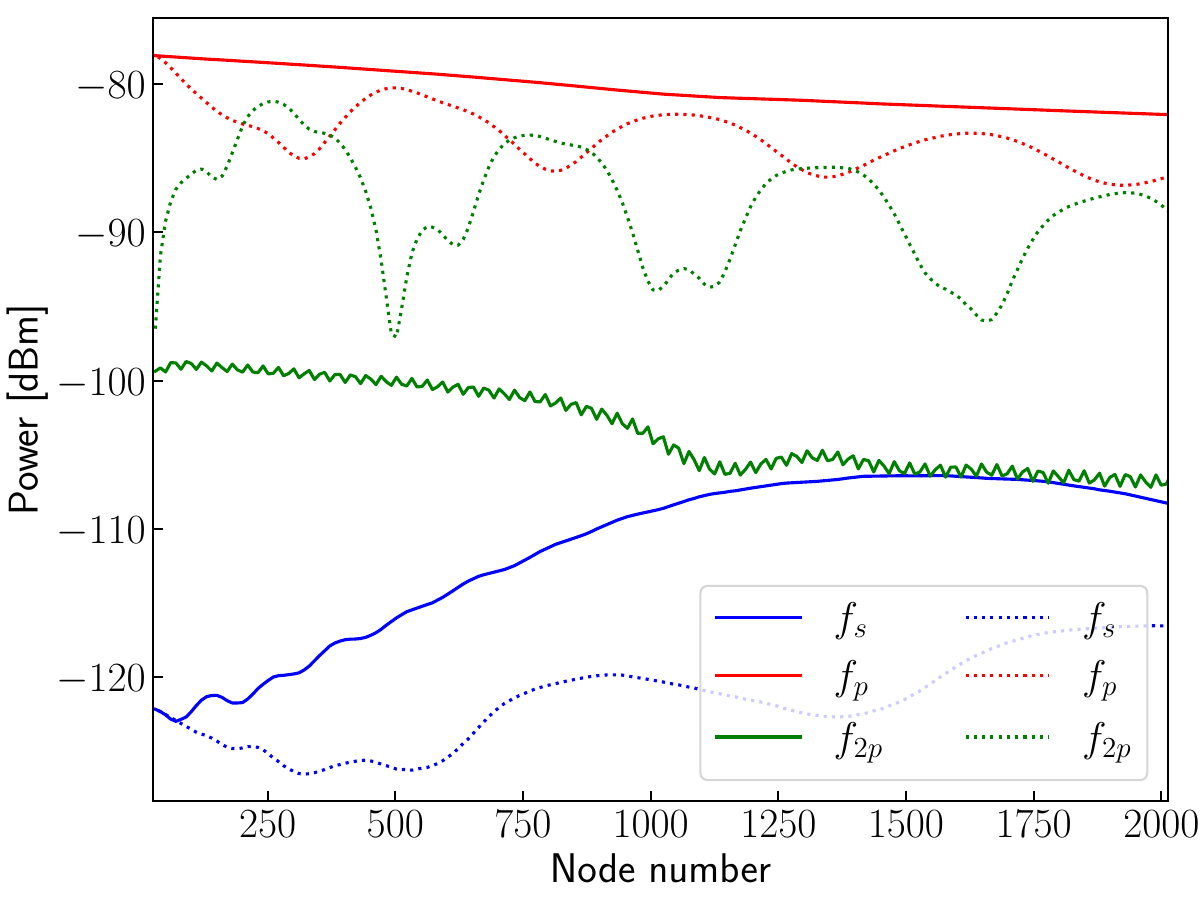}

 \caption{Power flow of propagating plane waves for main harmonics with plasma cutoff (solid lines) and without (dotted lines). Parameters of the 
simulation are: $f_{s} = 5 \textrm{ GHz}, f_{p} = 8.64 \textrm{ GHz}, I_{p} = 1.6\textrm{ uA}, I_{s} = 0.01\textrm{ uA}, I_{d} = 0.8\textrm{ uA}.$ Without plasma cutt-off it's clear that $f_{2p}$ harmonic took some power from the pump.
}
 \label{fig:Power_propagation}
\end{figure}

The agreement with CME is a noteworthy result since the theory does not account for higher harmonics, whose presence reduces the gain\cite{Dixon}. Also, the CME theory works within a constant pump approximation whereas the pump along the simulated TWPA slightly decreases. We believe that higher harmonics do not play a significant role in performance of proposed design, as they are suppressed by the plasma cut-off, as shown in Fig. \ref{fig:reflect} and demonstrated in Fig. \ref{fig:Power_propagation}.  Additionally, along with the high gain, the gain ripples are within approximately ~0.1 dB, which is negligible compared to, for instance, simulations of designs with periodic variations in the ground capacitance\cite{Gaydamachenko2022}.
 The signal and idler tones increase in an almost exponential manner with the devices length up to $1000$ Josephson junctions (see  Fig. \ref{fig:Power_propagation}). This suggests that up to this length, phase-matching is achieved (see Eq. (\ref{eq:gain})). Then, phase-mismatch develops and the optimal length is reached at the $1500$ Josephson junction. There are several reasons for phase mismatch. For example, the phase mismatch depends on the pump amplitude given in Eq. (\ref{eq:b}). The pump amplitude decreases as it's energy is transferred to other harmonics (see Fig. \ref{fig:Power_propagation}). Therefore, the phase-matching conditions changes with increasing amplifier length. 

Additionally to the phase matching, the transmission cutoff at the plasma frequency prevents the propagation of the pump's second harmonic. The oscillations at the frequency $2f_p$ are indeed generated, but they do not propagate. 
The comparison to the homogeneous case, i.e. without any dispersion engineering, is in Fig. \ref{fig:Power_propagation}.  Note that amplifiers with bandgaps produced by periodic modulation of their parameters \cite{Gaydamachenko2022,Planat2019, Nilsom_Photonic_crystall} exhibit the same feature. Indeed, the calculation shows that a low plasma frequency creates a cutoff which plays similar role as a bandgap and frequencies above $17.5$ \textrm{GHz} are  completely reflected (see Fig. \ref{fig:reflect}). Lower-frequency harmonics are less reflected  because they are farther from the frequency cutoff. 

\section{\label{sec:con}Conclusion}
To conclude, we have introduced a new TWPA design with phase matching  provided by  resonators created by the Josephson junction shunted with additional capacitor. Moreover, this design ensures the attenuation of higher harmonics. Previous works have created several band gaps using photonic crystals \cite{Planat2019} or by periodically modulating parameters \cite{Nilsom_Photonic_crystall, Kissling2023, Gaydamachenko2022}. Here, we demonstrated that creating just one band gap is sufficient, resulting in  simpler amplifier design. It only requires to increase the capacitance of every 5th junction, which is feasible to fabricate.

The modified theory of TL with resonant elements has been developed to calculate the dispersion relation when a resonator is placed at every nth node. This theory was validated using simulations in JoSIM and WRspice software. Despite the amplifier showing excellent results in both programs, the results were slightly different.

\begin{acknowledgments}
The work was supported in part by No. APVV-20-0425
SAS-MVTS, Grant QuantERA-SiUCs,
SPS Programme NATO grant number G5796, the Comenius University in Bratislava CLARA@UNIBA.SK high performance computing facilities, services and staff expertise of Centre for Information Technology.
The authors would like to thank Roman Martoňák and Matej Badin, for their assistance in setting up the simulator
platform and sharing their numerical cluster.
\end{acknowledgments}

\section*{Data Availability Statement}

The data that support the findings of
this study are available from the
corresponding author upon reasonable
request.

\appendix

\section{\label{app:elem} Model for periodic modifications to TL}

Rewriting Eq.(\ref{eq:Kirchhoff_law}) in the matrix form to the first order in $dz$ yields
\begin{equation}
    \begin{pmatrix}
        V(z+dz)\\
        I(z+dz) 
    \end{pmatrix}
    =
    M
    \begin{pmatrix}
        V(z)\\
        I(z) 
    \end{pmatrix},
\end{equation}
where
\begin{equation}
M=
    \begin{pmatrix}
        1 & -Z_Sdz\\
        -Y_Gdz & 1 
    \end{pmatrix}
\end{equation}
is the inverse ABCD matrix for the single TL element.
In Fig. \ref{fig:Scheme}e, a different element (with  series impedance $Z_S^\prime(\omega)$ and ground admittance $Y_G\prime(\omega)$) is inserted after every sequence of
$n$ identical elements ($Z_S^\prime(\omega)$, $Y_G^\prime(\omega)$). This cascade can be described by the following inverse ABCD matrix.
\begin{equation}
    \begin{pmatrix}
        1 & -Z_Sdz\\
        -Y_Gdz & 1 
    \end{pmatrix}^n
    \begin{pmatrix}
        1 & -Z_S^\prime dz\\
        -Y_G^\prime dz & 1 
    \end{pmatrix}
\end{equation}
The nth power of the "not-primed" matrix $M$ can be calculated by diagonalization. The  nth power of eigenvalues of matrix $M$ are to the first order in $dz$ 
\begin{equation}
    \lambda_{\pm}^n = (1\pm\sqrt{ZY}dz)^n \approx 1\pm n\sqrt{ZY}dz,
\end{equation}
which, together with the "primed" matrix, yields the final matrix of $n+1$ elements 
\begin{equation}
    M_{n+1}
    \begin{pmatrix}
        1 & -(nZ_S+Z_S^\prime)dz\\
        -(nY_G+Y_G^\prime)dz & 1 
    \end{pmatrix}
\end{equation}
The matrix $M_{n+1}$ describes an element with series impedance $(nZ_S+Z_S^\prime)dz$ and ground admittance $(nY_G+Y_G^\prime)dz$ which justifies the approximation in Eq. (\ref{eq:plas_dis}).

\section{\label{app:comp} JoSIM and WRspice comparison}

\label{app:progs}

\begin{figure}[h]

\includegraphics[width=8.6cm]{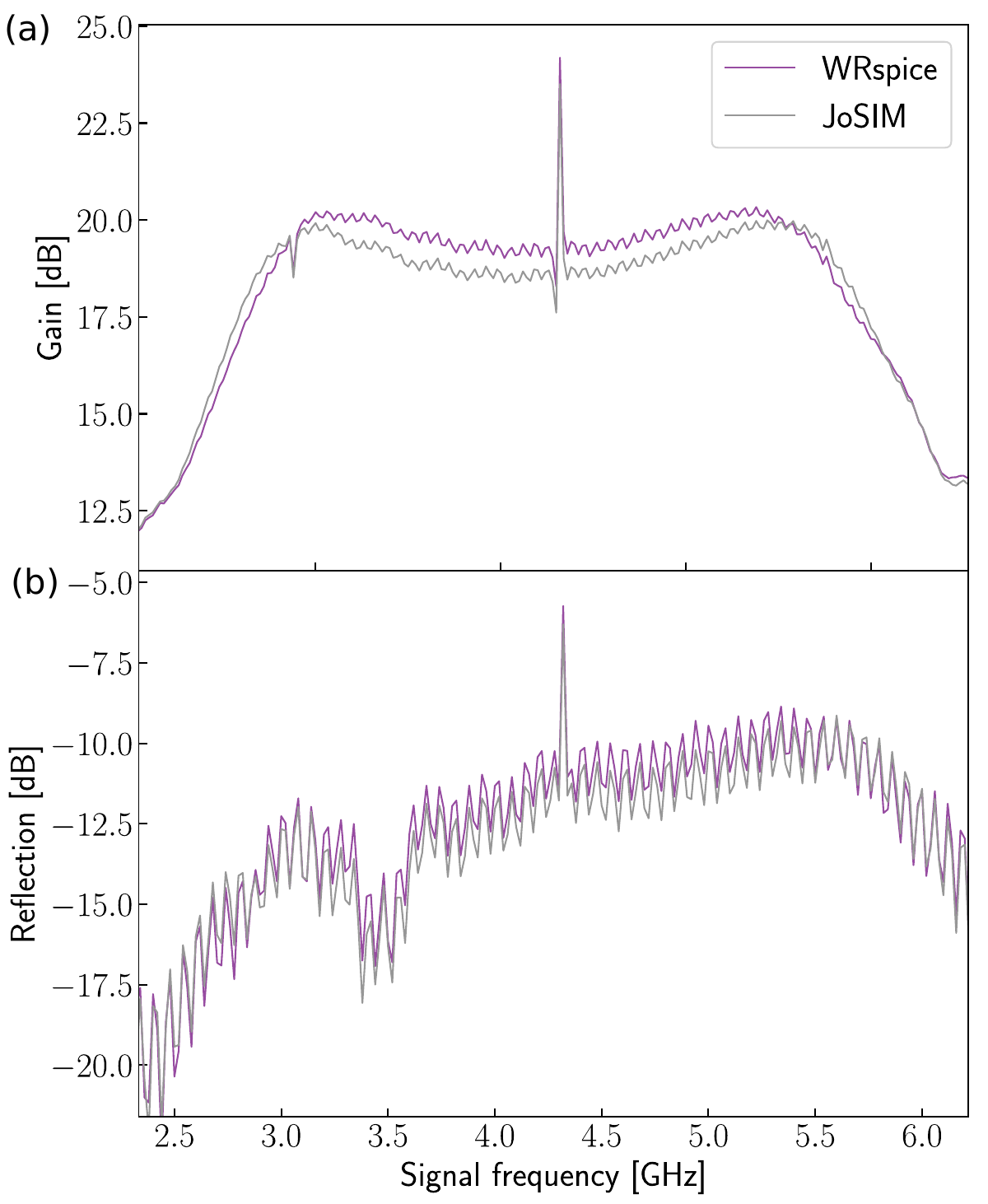}

\caption{\label{fig:Gain_node_comparing}Gain (a) and reflection (b) simulated in JoSIM (purple) and WRspice (grey) for  $I_{s} = 0.01\textrm{ uA}$, $I_{dc} =0.8\textrm{ uA}, f_{p}=8.64\textrm{ GHz}$. Pump amplitudes are different in JoSIM ($I_{p} = 1.6\textrm{ uA}$) and WRspice ($I_{p} = 1.65\textrm{ uA}$). }
\end{figure}

At the optimal working point the gain profile and gain amplitude in JoSIM and WRspice (maximum amplitude at reflection less than $-10 \textrm{ dB}$) are approximately the same Fig. \ref{fig:Gain_node_comparing}. However, the WRspice requires higher pumping amplitude. JoSIM simulates noticeably faster, so we recommend using it first and then verifying the results with WRspice.

%\nocite{*}
\bibliographystyle{apsrev}

\bibliography{main}% Produces the bibliography via BibTeX.

\end{document}